\documentclass[pra,twocolumn,showpacs,floatfix]{revtex4-1}
\usepackage{amssymb}
\usepackage{graphicx}
\usepackage[english]{babel}
\usepackage{latexsym}
\usepackage{graphics}
\usepackage{subfigure}

\def\be{\begin{equation}}
\def\ee{\end{equation}}
\def\bea{\begin{eqnarray}}
\def\eea{\end{eqnarray}}
\def\bse{\begin{subequations}}
\def\ese{\end{subequations}}

\def\be{\begin{eqnarray}}
\def\ee{\end{eqnarray}}

\begin{document}

\title{Geometric optics of Bloch waves in a chiral and dissipative medium}
\author{Chuanwei Zhang$^{1}$, and Qian Niu$^{2}$}

\begin{abstract}
We present a geometric optics theory for the transport of quantum particles
(or classical waves) in a chiral and dissipative periodic crystal subject to
slowly varying perturbations in space and time. Taking account of some
properties of particles and media neglected in previous theory, we find
important additional terms in the equations of motion of particles. The
(energy) current density field, which traces the geometric optics rays, is
not only governed by the Bloch band energy dispersion but also involves
there additional fields. These are the angular momentum of the particle, the
dissipation dipole density, and various geometric gauge fields in the
extended phase space spanned by space-time and its reciprocal, momentum and
frequency. For simplicity, the theory is presented using light propagation
in photonic crystals.
\end{abstract}

\pacs{03.65.Sq, 42.25Bs, 03.65.Vf, 72.10.-d}
\maketitle

\affiliation{$^{1}$Department of Physics and Astronomy, Washington State University,
Pullman, Washington, 99164 USA\\
$^{2}$Department of Physics, The University of Texas, Austin, Texas 78712 USA}

\section{Introduction}

The transport property of quantum particles (or classical waves) in media is
one of the central problems in physics. A textbook example is the dynamics
of Bloch electrons in solid state crystals \cite{Mermin,Marder}. For a long
time, the semiclassical equations of motion (EOM) of electrons are known to
be solely determined by the energy band dispersion \cite{Mermin}. Recently
it was found that geometric gauge fields due to Berry phase \cite{Berry}
effects may modify the EOM of particles \cite{Marder,Niu1} in a chiral
medium, leading to new and important physical phenomena and applications
\cite{Xiao1,Hall,Hall2,onoda,bliokh2,Haldane,Nagaosa1,Yang}. Here
\textquotedblleft chiral" means that the time reversal or spatial inversion
symmetry in the medium is broken.

However, the EOM in current literature neglect several important properties
of particles (or waves) and media (\textit{e.g.}, self rotation of
particles, frequency (or energy) dependence and dissipation of the media).
In this paper, we derive a new set of EOM of particles where the effects of
these properties are included. For simplicity of the presentation, we
consider a concrete physical system: light propagation in photonic crystals
\cite{Joan97}, although the results are applicable to many other systems
\cite{Nagaosa1}. Instead of wavepacket dynamics commonly used in previous
literature, we adopt a geometric optics approach that has been proved to be
extremely valuable for describing light propagation in usual optical
materials. However, the traditional approach of geometric optics is
inadequate because the light wavelength is typically the same order as the
unit cell size of the photonic crystals within which the dielectric constant
varies.

The new geometric optics theory is presented for light propagation in a
frequency band of a chiral and dissipative photonic crystal under
perturbations which vary slowly in space and time compared to the lattice
constant and the band gaps. We focus on the Poynting vector field, averaged
over the crystal unit cell, which traces the geometric optics rays (\textit{%
i.e.}, the EOM). It involves the photon wavepacket center velocity, and also
contains two additional terms due to the photon orbital angular momentum
(POAM) and the dissipation of the medium that are absent in previous
wavepacket theory of photons \cite{onoda,bliokh2}. We show that, in a
concrete example, the term from POAM can dramatically modify the transport
properties of light in photonic crystals. The propagation of light is also
affected by various geometric gauge fields due to Berry phase effects. Both
POAM and Berry phase originate from the chirality of the medium or the
particle. Two such gauge fields mimic closely the electric and magnetic
fields for electrons \cite{Yang}, and two other fields arise from Berry
curvatures in the momentum-frequency space, previously called reciprocal
electromagnetic fields in the context of electron semiclassical dynamics
\cite{Niu1,Shindou}. The physical origin of the reciprocal electric field is
illustrated. There are also fields from Berry curvatures in other facets of
the extended phase space. All these fields can be calculated based on the
band structure wave functions of the photonic crystal.

The paper is organized as follows: Section II introduces the wave equations
and the wavepacket dynamics for light propagation in photonic crystals. In
section III, a geometric optics theory for light propagation in photonic
crystals is developed. Section IV gives an example of the application of the
theory. Finally, we summarize and conclude in Section V.

\section{Wave equations and wavepacket dynamics:}

Consider light propagating in an inhomogeneous photonic crystal
characterized by the dielectric permittivity tensor $\mathbf{\varepsilon }%
\left( \mathbf{r}\right) $ and magnetic permeability tensor $\mathbf{\mu }%
\left( \mathbf{r}\right) $, where $\mathbf{\varepsilon }$ and $\mathbf{\mu }$
are spatially fast-varying periodic functions with slowly spatial
modulation. $\mathbf{\varepsilon }$ and $\mathbf{\mu }$ may also have smooth
temporal modulation, which, together with dissipation of the medium, will be
discussed in Sec. III.D for simplicity of the presentation. The Maxwell
equations without external sources (\textit{i.e.}, currents) can be written
as a Schr\"{o}dinger-equation like form
\begin{equation}
\left[ \mathcal{H}\left( -i\partial _{\mathbf{r}},\mathbf{r}\right) -\Xi
\left( \omega ,\mathbf{r}\right) \right] \Phi \left( \mathbf{r}\right) =0,
\label{waveeq}
\end{equation}%
where $\mathcal{H}=\left(
\begin{array}{cc}
0 & i\partial _{\mathbf{r}}\times \\
-i\partial _{\mathbf{r}}\times & 0%
\end{array}%
\right) $, $\Phi =\left( \mathbf{E,H}\right) ^{T}/\sqrt{2}$ is the
electromagnetic wavefunction, $\Xi \left( \omega \right) =\omega \left(
\begin{array}{cc}
\mathbf{\varepsilon }\left( \omega \right) & 0 \\
0 & \mathbf{\mu }\left( \omega \right)%
\end{array}%
\right) $ and the frequency $\omega $ corresponds to $i\partial _{t}$ in the
time-dependent Maxwell equation. In multiferroic materials with strong
magnetoelectric coupling, $\Xi \left( \omega \right) $ may also have
non-diagonal terms \cite{Cheong}. The form of the wave equation (\ref{waveeq}%
) is very general. For instance, it can be taken as the Schr\"{o}dinger
equation for $\Xi =\hbar \omega $ and $\mathcal{H}=-\hbar ^{2}\partial _{%
\mathbf{r}}^{2}/2m+V\left( \mathbf{r}\right) $.

Without considering dissipation and frequency dependence of the medium, the
wave equation (\ref{waveeq}) can be treated using the wavepacket dynamics.
The central idea \cite{Niu1} is to construct a wavepacket
\begin{equation}
\left\vert \Phi \right\rangle =\sqrt{\rho }\int d^{3}qa\left( \mathbf{q}%
,t\right) \left\vert \psi \left( \mathbf{q},\mathbf{r}_{c}\right)
\right\rangle
\end{equation}%
with a mean wavevector
\begin{equation}
\mathbf{q}_{c}=\int d^{3}q\left\vert a\left( \mathbf{q},t\right) \right\vert
^{2}\mathbf{q}
\end{equation}%
and a preassigned wavepacket center position
\begin{equation}
\mathbf{r}_{c}=[\left\langle \Phi \right\vert \mathbf{\hat{r}}\left\vert
\Theta _{0}\Phi \right\rangle +\left\langle \Theta _{0}\Phi \right\vert
\mathbf{\hat{r}}\left\vert \Phi \right\rangle ]/2\rho .
\end{equation}%
Here $\psi \left( \mathbf{q},\mathbf{r}_{c}\right) =\left( \mathbf{E,H}%
\right) ^{T}/\sqrt{2\rho }=e^{i\mathbf{q}\cdot \mathbf{\hat{r}}}\phi \left(
\mathbf{q},\mathbf{r}_{c}\right) $ is the Bloch eigenstate of the local wave
equation
\begin{equation}
\lbrack \mathcal{H}_{0}-\Xi _{0}\left( \omega _{c0},\mathbf{r}_{c}\right)
]\psi \left( \mathbf{q},\mathbf{r}_{c}\right) =0  \label{waveq1}
\end{equation}%
with the eigenenergy $\omega _{c0}=\omega _{c0}\left( \mathbf{q},\mathbf{r}%
_{c}\right) $, $\mathcal{H}_{0}-\Xi _{0}$ is the local operator with the
required periodicity of the unperturbed crystal, $\Theta _{0}=\partial \Xi
_{0}/\partial \omega _{c0}$, $\phi \left( \mathbf{q},\mathbf{r}_{c}\right) $
is the periodic part of the Bloch wave and satisfies $\left\langle \phi
\right\vert \Theta _{0}\left\vert \phi \right\rangle =1$, $\rho
=\left\langle \Phi \right\vert \Theta _{0}\left\vert \Phi \right\rangle $ is
the normalization factor. In a medium without frequency dependence, we have $%
\rho =\left( \mathbf{E}\cdot \mathbf{D}+\mathbf{H\cdot B}\right) /2$, the
energy density of photons. While in the Schr\"{o}dinger equation, $\rho $ is
the wavedensity. The amplitude $a\left( \mathbf{q},t\right) $ satisfy the
normalization condition $\int d^{3}q\left\vert a\left( \mathbf{q},t\right)
\right\vert ^{2}=1$.

Using a time-dependent variational principle with the Lagrangian
\begin{equation}
L=\left\langle \Phi \right\vert \left[ \Xi \left( i\partial _{t},\mathbf{r}%
\right) -\mathcal{H}\left( -i\partial _{\mathbf{r}},\mathbf{r}\right) \right]
\left\vert \Phi \right\rangle ,
\end{equation}%
we obtain the semiclassical EOM for photons
\begin{eqnarray}
\mathbf{\dot{r}}_{c} &=&\frac{\partial \omega _{c}}{\partial \mathbf{q}_{c}}%
-\Omega _{\mathbf{qq}}\mathbf{\dot{q}}_{c}-\Omega _{\mathbf{qr}}\mathbf{\dot{%
r}}_{c},  \label{vel} \\
\mathbf{\dot{q}}_{c} &\mathbf{=}&-\frac{\partial \omega _{c}}{\partial
\mathbf{r}_{c}}+\Omega _{\mathbf{rq}}\mathbf{\dot{q}}_{c}+\Omega _{\mathbf{rr%
}}\mathbf{\dot{r}}_{c}  \label{mom}
\end{eqnarray}%
by following a similar derivation as that in Ref. \cite{Niu1} for electrons,
where
\begin{equation}
\omega _{c}=\omega _{c0}+\text{Im}\left\langle \frac{\partial \phi }{%
\partial \mathbf{r}_{c}}\right\vert \cdot \left( \mathcal{H}_{0}-\Xi
_{0}\right) \left\vert \frac{\partial \phi }{\partial \mathbf{q}_{c}}%
\right\rangle
\end{equation}%
is the total energy,
\begin{equation}
\Omega _{\mathbf{qq}}^{\alpha \beta }=i\left\langle \frac{\partial \phi }{%
\partial \mathbf{q}_{\alpha }}\right\vert \Theta _{0}\left\vert \frac{%
\partial \phi }{\partial \mathbf{q}_{\beta }}\right\rangle +c.c.  \label{BC}
\end{equation}%
is the Berry curvature in the momentum space, $\Omega _{\mathbf{qr}}$ and $%
\Omega _{\mathbf{rr}}$ have similar definitions.

\section{Geometric optics approach}

The geometric optics approximation we adopt does not involve the concept of
wavepacket. Substituting the WKB trial wavefunction $\Phi \left( \mathbf{r}%
,t\right) =\exp \left[ iS\left( \mathbf{r},t\right) \right] u\left( \mathbf{r%
}\right) $ into the time-dependent Maxwell equation, we obtain
\begin{equation}
\left[ \mathcal{H}\left( -i\partial _{\mathbf{r}}+\mathbf{p}\right) -\Xi
\left( \omega ,\mathbf{r}\right) \right] u=0,  \label{waveeq3}
\end{equation}%
where $\mathbf{p}=\partial _{\mathbf{r}}S$ is the wavevector, the frequency $%
\omega =-\partial _{t}S$ is the same as that defined in Eq. (\ref{waveeq}). $%
\mathbf{p}$ plays a similar role as $\mathbf{q}_{c}$ in the wavepacket
dynamics. In the geometric optics theory for an inhomogeneous medium, $%
\mathbf{p}$'s direction is perpendicular to the wave front that has a
constant phase $S$.

Because the spatial modulation of the medium is slowly varying, $u$ can be
taken as a periodic function of $\mathbf{r}$ over a few lattice sites. The
spatial modulation of $u$ can be accounted by taking $u$ as a function of a
slowly varying variable $\mathbf{R}$, where $\mathbf{R}$ describes
variations on a length scale much larger than the lattice spacing. $\mathbf{R%
}$ plays the same role as $\mathbf{r}_{c}$. For a fixed $\mathbf{R}$, $u$ is
a periodic function of the fast varying variable $\mathbf{\bar{r}}$. With
the slowly varying variable $\mathbf{R}$ and the fast varying variable $%
\mathbf{\bar{r}}$, Eq. (\ref{waveeq3}) can be rewritten as
\begin{equation}
\left[ \mathcal{H}\left( -i\partial _{\mathbf{\bar{r}}}-i\partial _{\mathbf{R%
}}+\mathbf{p}\right) -\Xi \left( \omega ,\mathbf{\bar{r}},\mathbf{R}\right) %
\right] u=0.  \label{waveeq2}
\end{equation}%
By choosing a suitable phase $S$, we can take $\mathbf{p}$ and $\omega $ as
functions of $\mathbf{R}$.

Because $u$ varies slowly with respect to $\mathbf{R}$, we may solve Eq. (%
\ref{waveeq2}) perturbatively. Up to the first order correction $\partial _{%
\mathbf{R}}u$, we can take the gradient expansion $\mathcal{H}\left(
-i\partial _{\mathbf{\bar{r}}}-i\partial _{\mathbf{R}}+\mathbf{p}\right)
\approx \mathcal{H}_{0}\left( -i\partial _{\mathbf{\bar{r}}}+\mathbf{p}%
\right) +\Delta \mathcal{H}$ with
\begin{equation}
\Delta \mathcal{H}=-\frac{i}{2}\left( \partial _{\mathbf{R}}\cdot \frac{%
\partial \mathcal{H}_{0}}{\partial \mathbf{p}}+\frac{\partial \mathcal{H}_{0}%
}{\partial \mathbf{p}}\cdot \partial _{\mathbf{R}}\right) ,
\end{equation}%
and $\Xi \left( \omega ,\mathbf{\bar{r},R}\right) =\Xi _{0}\left( \omega
_{0}\right) +\Theta _{0}\omega _{1}$. Here we denote
\begin{equation}
\Theta _{0}=\frac{\partial \Xi _{0}\left( \omega _{0}\right) }{\partial
\omega _{0}}
\end{equation}%
and expand the eigenenergy and wavefunction to the first order: $\omega
=\omega _{0}+\omega _{1}$, $u=u_{0}+u_{1}$. With these expansions, the zero
order of Eq. (\ref{waveeq2}) gives%
\begin{equation}
\left[ \mathcal{H}_{0}\left( -i\partial _{\mathbf{\bar{r}}}+\mathbf{p}%
\right) -\Xi _{0}\right] u_{0}=0,  \label{zero}
\end{equation}%
which is similar as the local wave equation (\ref{waveq1}) in the wavepacket
dynamics. For a fixed $\mathbf{R}$, Eq. (\ref{zero}) yields a Bloch
wavefunction $u_{0}=\sqrt{\rho }f$, where we choose the normalization
condition $\left\langle f\right\vert \Theta _{0}\left\vert f\right\rangle
=\int d^{3}\mathbf{\bar{r}}f^{\ast }\left( \mathbf{\bar{r}}\right) \Theta
_{0}f\left( \mathbf{\bar{r}}\right) =1$. Here and later in this article, an
inner product corresponds to an integration over a unit lattice cell. $f$
plays the same role as $\phi $ in the wavepacket dynamics. We emphasize that
the wavefunction $f$ is a periodic function of the fast varying variable $%
\mathbf{\bar{r}}$ that originates from the periodicity of the photonic
crystals. $f$ describes the periodic wave property of light in photonic
crystals that does not exist in traditional geometric optical approach for
an inhomogeneous medium. The periodic wavefunction can lead to Eq. (\ref%
{zero}), together with suitable boundary conditions, yields the energy
dispersion relation $\omega _{0}=\omega _{0}\left( \mathbf{p},\mathbf{R}%
\right) $. In general $f$ may be a function of $\mathbf{\bar{r}}$, $\mathbf{R%
}$, $\mathbf{p}$ and $\omega _{0}$, depending on the way to record the
wavefunction. In this article, we will take $f$ as a function of $\mathbf{%
\bar{r}}$, $\mathbf{p}$, $\mathbf{R}$ only by replacing $\omega _{0}$ in $%
u_{0}$ as $\omega _{0}\left( \mathbf{p},\mathbf{R}\right) $ using the
dispersion relation.

The first order of the gradient expansion of Eq. (\ref{waveeq2}) is%
\begin{equation}
\mathcal{L}_{0}u_{1}=\mathcal{G}u_{0},  \label{1stcor}
\end{equation}%
where $\mathcal{L}_{0}=\mathcal{H}_{0}-\Xi _{0}$, $\mathcal{G}=\mathcal{-}%
\Delta \mathcal{H}+\Theta _{0}\omega _{1}$. Note that the semiclassical
geometric optics approach is different from the well-known Luttinger-Kohn
treatment \cite{Luttinger}, which is a full quantum mechanical perturbative
approach to perturbed periodic systems. In Luttinger-Kohn treatment, the
wavefunction is expanded around the band minimum. While in the geometric
optics approach, the gradient expansion is around the local Bloch
wavefunction.

\subsection{Energy correction and Berry phase}

Multiplying each side of Eq. (\ref{1stcor}) with $\left\langle
u_{0}\right\vert $, we obtain the first order energy correction $\omega
_{1}=\left\langle f\right\vert \Delta \mathcal{H}\left\vert f\right\rangle $%
. A straightforward but tedious evaluation of $\omega _{1}$ yields
\begin{equation}
\omega _{1}=\text{Im}\left\langle \frac{\partial f}{\partial \mathbf{R}}%
\right\vert \cdot \mathcal{L}_{0}\left\vert \frac{\partial f}{\partial
\mathbf{p}}\right\rangle -\text{Im}\left\langle f\right\vert \Theta
_{0}\left\vert \frac{df}{dt}\right\rangle .  \label{enecor2}
\end{equation}%
The first term $\omega _{1a}=\text{Im}\left\langle \frac{\partial f}{%
\partial \mathbf{R}}\right\vert \cdot \mathcal{L}_{0}\left\vert \frac{%
\partial f}{\partial \mathbf{p}}\right\rangle $ is the same as that derived
from the wavepacket dynamics. In the second term $\omega _{1b}=-\text{ Im}%
\left\langle f\right\vert \Theta _{0}\left\vert \frac{df}{dt}\right\rangle $%
, $\frac{df}{dt}=\mathbf{\dot{R}}\cdot \frac{\partial f}{\partial \mathbf{R}}%
+\mathbf{\dot{p}}\cdot \frac{\partial f}{\partial \mathbf{p}}$ since we
assume no explicit time-dependence of the medium. Therefore along a path $%
\mathcal{C}$, a Berry phase $\gamma =\int_{\mathcal{C}}\mathbf{A}_{\mathbf{R}%
}\cdot d\mathbf{R}+\mathbf{A}_{\mathbf{p}}\cdot d\mathbf{p}$ is accumulated.
Here $\mathbf{A}_{\mathbf{R}}$ and $\mathbf{A}_{\mathbf{p}}$ are Berry
connections in position and momentum spaces, with $\mathbf{A}_{\mathbf{R}}=-%
\text{Im}\left\langle f\right\vert \Theta _{0}\left\vert \frac{\partial f}{%
\partial \mathbf{R}}\right\rangle $. $\mathbf{A}_{\mathbf{p}}$ has a similar
definition.

The procedure to obtain the energy correction (\ref{enecor2}) also yields
the continuity equation
\begin{equation}
\frac{\partial \rho }{\partial t}+\partial _{\mathbf{R}}\cdot \mathbf{J}=0,
\label{conteq}
\end{equation}
where
\begin{equation}
\mathbf{J}=\left\langle u\right\vert \frac{\partial \mathcal{H}}{\partial
\mathbf{p}}\left\vert u\right\rangle =\mathbf{E}\times \mathbf{H}
\label{current}
\end{equation}%
is the Poynting vector (the energy current density) of light. Note that $%
\rho $ does not vary with time without considering the dissipation of the
medium, \textit{i.e.}, $\frac{\partial \rho }{\partial t}=0$. At the zero
order, $\mathbf{J}_{0}$ $=\rho \mathbf{v}_{0}$ with $\mathbf{v}_{0}=\partial
\omega _{0}/\partial \mathbf{p}$ as the zero order of the local velocity. In
the Schr\"{o}dinger equation, $\mathbf{J}$ is the current density of
particles.

\subsection{Poynting vector field}

Substituting $u=u_{0}+u_{1}$ into (\ref{current}), we find the Poynting
vector can be rewritten as
\begin{equation}
\mathbf{J}=\rho \left( \frac{\partial \bar{\omega}}{\partial \mathbf{p}}%
-\Omega _{\mathbf{pp}}\mathbf{\dot{p}}-\Omega _{\mathbf{pR}}\mathbf{v}%
_{0}\right) +\partial _{\mathbf{R}}\times \rho \mathbf{m}.  \label{current3}
\end{equation}%
Here $\bar{\omega}=\omega _{0}+\omega _{1a}$, $\mathbf{\dot{p}=v}_{0}\cdot
\frac{\partial \mathbf{p}}{\partial \mathbf{R}}$, $\partial _{\mathbf{R}}=%
\frac{\partial }{\partial \mathbf{R}}+\frac{\partial \mathbf{p}}{\partial
\mathbf{R}}\cdot \frac{\partial }{\partial \mathbf{p}}$, $\mathbf{m}$ is the
POAM defined in Eq. (\ref{Orbit}). The Berry curvatures $\Omega _{\mathbf{pp}%
}$ and $\Omega _{\mathbf{pR}}$ have the same definitions as that in the
wavepacket dynamics (Eq. (\ref{BC})) with the replacement of the notations $%
\left\{ \mathbf{r}_{c},\mathbf{q}_{c},\omega _{c0},\phi \right\} \rightarrow
\left\{ \mathbf{R},\mathbf{p},\omega _{0},f\right\} $. The first three terms
in Eq. (\ref{current3}) describe the energy flow due to the translational
motion of photons with a velocity same as Eq. (\ref{vel}) for the wavepacket
dynamics.

It is clear from Eq. (\ref{current3}) that the energy current density $%
\mathbf{J}$ cannot be simply determined by the translational motion of
photons. $\mathbf{J}$ contains an additional term $\partial _{\mathbf{R}%
}\times \rho \mathbf{m}$, which is absent in Eq. (\ref{vel}) for the
wavepacket dynamics. Note that
\begin{equation}
\mathbf{m=}-\frac{i}{2}\left\langle \frac{\partial f}{\partial \mathbf{p}}%
\right\vert \times \left( \Xi _{0}-\mathcal{H}_{0}\right) \left\vert \frac{%
\partial f}{\partial \mathbf{p}}\right\rangle  \label{Orbit}
\end{equation}%
is the POAM. For non-interacting electrons in crystals (no frequency
dependence), $\mathbf{m}$ is just the orbital magnetic moment of Bloch
electrons \cite{Niu1}. In the wave-packet treatment of electron dynamics, $%
\mathbf{m}$ originates from the finite spread of the wavepacket. The
wavepacket generally rotates about its center position, giving rise to an
orbital magnetic moment $\mathbf{m}=-e/2\left\langle W\right\vert \left(
\mathbf{r}-\mathbf{r}_{c}\right) \times \mathbf{\hat{v}}\left\vert
W\right\rangle $, where $\left\vert W\right\rangle $ is the wave packet and $%
\mathbf{\hat{v}}$ is the velocity operator \cite{Xiao}. Therefore the term $%
\partial _{\mathbf{R}}\times \rho \mathbf{m}$ corresponds to a local current
density field arising from the self-rotation of photons (\textit{i.e.},
POAM) in photonic crystals.

Note that both $\mathbf{p=\partial }_{\mathbf{r}}S$, and $\omega =-\partial
_{t}S$ are gauge dependent quantities, \textit{i.e.}, they depend on the
choice of $S$. In the static case, a natural choice of the gauge is to
ensure that energy is the same for different positions, \textit{i.e.}, $%
\partial _{\mathbf{R}}\omega =0$. Under this gauge, we find $\mathbf{\dot{p}=%
}-\frac{\partial \omega _{0}}{\partial \mathbf{R}}$, which is the same as
Eq. (\ref{mom}) in the wavepacket dynamics up to the first order of the
gradient expansion. It is easy to check all terms in Eq. (\ref{current3})
are gauge independent. Therefore the Poynting vector $\mathbf{J}$ does not
depend the choice of $S$. The equations for $\mathbf{J}=\rho \mathbf{\dot{R}}
$ and $\mathbf{\dot{p}}$ determine the geometric ray self-consistently. With
these two equations, we can obtain the position and momentum parameters $%
\mathbf{R}$, $\mathbf{p}$ at any time.

The gauge invariant momentum and energy defined in the wavepacket dynamics
can be related with $\mathbf{p}$, $\omega $ through $\mathbf{q}_{c}=$ $%
\mathbf{p-}$ $\mathbf{A}_{\mathbf{R}}$, $\omega _{c}=\omega _{0}+\omega _{1}+%
\text{Im}\left\langle f\right\vert \Theta _{0}\left\vert \frac{df}{dt}%
\right\rangle $. Under this transformation, we can show that the equations
for $\mathbf{\dot{R}}$ and $\mathbf{\dot{p}}$ transfer to Eqs. (\ref{vel})
and (\ref{mom}) derived from the wavepacket dynamics (except the term $%
\partial _{\mathbf{R}}\times \rho \mathbf{m}$). Note that the term $\Omega _{%
\mathbf{RR}}\mathbf{\dot{R}}$\ in Eq. (\ref{mom}) (with $\mathbf{R}$
replacing $\mathbf{r}_{c}$) can be rewritten as $\mathbf{\dot{R}}\times
\mathcal{B}$, where $\mathcal{B}=\frac{1}{2}\epsilon _{\alpha \beta \gamma
}\Omega _{\mathbf{RR}}^{\beta \gamma }$ is an \textit{effective magnetic
field} for photons.

\subsection{Reciprocal electric field}

It is well known that the term $-\Omega _{\mathbf{pp}}\mathbf{\dot{p}}$ in
the equation of motion for $\mathbf{\dot{R}}$ can be rewritten as $-\mathbf{%
\dot{p}}\times \mathbf{\Omega }$ with $\mathbf{\Omega }=\frac{1}{2}\epsilon
_{\alpha \beta \gamma }\Omega _{\mathbf{pp}}^{\beta \gamma }$ as the \textit{%
reciprocal magnetic field} in the momentum space \cite{Niu1}. Interestingly,
a \textit{reciprocal electric field }may also exist in a frequency-dependent
medium. In a frequency-dependent medium \cite{Shvets}, if the wavefunction $%
f $ is chosen in a way that has explicit $\omega _{0}$ dependence, the
derivative $\frac{\partial }{\partial \chi }$ ($\chi $ can be $\mathbf{R}$, $%
\mathbf{p}$) becomes $\frac{\partial }{\partial \chi }+\frac{\partial \omega
_{0}}{\partial \chi }\frac{\partial }{\partial \omega _{0}}$ and many new
terms depending on the Berry curvature in the frequency space may appear in
the Poynting vector (\ref{current3}). For instance, the second term in Eq. (%
\ref{current3}) can be rewritten as
\begin{equation}
-\rho \mathbf{\dot{p}}\times \left[ \mathbf{\Omega }+\mathbf{v}_{0}\times
\Upsilon \right] ,
\end{equation}%
where%
\begin{equation}
\Upsilon =i\left\langle \frac{\partial f}{\partial \omega _{0}}\right\vert
\Theta _{0}\left\vert \frac{\partial f}{\partial \mathbf{p}}\right\rangle
+c.c  \label{REF}
\end{equation}%
is the \textit{reciprocal electric field} in the momentum-frequency space.
It is interesting that the form of $\Upsilon $ obtained from the simple
substitution here agrees with that obtained in Ref. \cite{Shindou} for
quasiparticle dynamics in an interacting Fermi liquid based on the
complicated Keldysh formalism, where the self-energy of the quasiparticles
has an energy dependence.

\subsection{Slow time modulation and dissipation of the medium}

In the presence of slow time modulation of $\mathbf{\varepsilon }$ and $%
\mathbf{\mu }$, characterized by a slow variable $T$ (similar as $\mathbf{R}$%
), additional terms $\Omega _{T\mathbf{q}}$ and $-\Omega _{T\mathbf{R}}$
appear in the equations of motion (\ref{vel}) (also in (\ref{current3})) and
(\ref{mom}) respectively. The term $-\Omega _{T\mathbf{R}}=\frac{\partial
\mathbf{A}_{\mathbf{R}}}{\partial T}-\frac{\partial A_{T}}{\partial \mathbf{R%
}}$ ($A_{T}$ is the Berry connection in the $T$ space) can be taken as an
\textit{effective electric field} for photons with $\left( -\mathbf{A}_{%
\mathbf{R}},A_{T}\right) $ serving as the gauge potential of the effective
magnetic and electric fields. Note that the effective magnetic field $%
\mathcal{B}$ and electric field $-\Omega _{T\mathbf{R}}$ are different from
the real magnetic and electric fields $(\mathbf{E},\mathbf{H})$ which are
contained in the wavefunctions of photons.

In previous discussion, we have neglected the dissipation of the medium for
simplicity of the presentation. We note that the dissipation of the medium
can be incorporated by adding an anti-Hermitian part $i\Gamma =\left( \Xi
-\Xi ^{\dag }\right) /2$ to the operator $\Xi =\bar{\Xi}+i\Gamma $, here $%
\Gamma $ is assumed to be a small perturbation to the Hermitian part $\bar{%
\Xi}=\left( \Xi +\Xi ^{\dag }\right) /2$. $\Theta _{0}$ in previous
equations is now replaced with $\Theta _{0}=\partial \bar{\Xi}_{0}/\partial
\omega $. The continuity equation (\ref{conteq}) becomes
\begin{equation}
\partial _{T}\rho +\nabla _{\mathbf{R}}\cdot \mathbf{J}=-\rho /\tau ,
\label{conti}
\end{equation}%
where $\tau ^{-1}=2\left\langle f\right\vert \Gamma \left\vert
f\right\rangle $ describes the dissipation of the medium. The right hand
side of the continuity equation (\ref{conti}) represents the change of the
energy density and Poynting vector field due to the dissipation of the
medium. For an isotropic medium ($\mathbf{\varepsilon }$ and $\mathbf{\mu }$
are scalar) with $\mathbf{H}=0$, we have%
\begin{equation}
\tau ^{-1}=2\omega _{0}\text{Im}\varepsilon \left( \omega _{0}\right) /\text{%
Re}\varepsilon \left( \omega _{0}\right) .  \label{diss2}
\end{equation}%
We see $\tau ^{-1}\text{Re}n/2$ corresponds to the attenuation constant or
absorption coefficient of a plane wave propagating in an absorptive medium
\cite{Jackson}, where $n$ is the index of refraction of the medium.

The dissipation of the medium yields an additional term to the Poynting
vector field (\ref{current3})%
\begin{equation}
\rho \digamma =\rho \text{Im}\left[ \left\langle f\right\vert \tau
^{-1}\Theta _{0}\left\vert \frac{\partial f}{\partial \mathbf{p}}%
\right\rangle -2\left\langle f\right\vert \Gamma \left\vert \frac{\partial f%
}{\partial \mathbf{p}}\right\rangle \right] ,  \label{diss}
\end{equation}%
which is also absent in Eq. (\ref{vel}) for the wavepacket dynamics. Clearly
this term is gauge invariant since $\tau ^{-1}=2\left\langle f\right\vert
\Gamma \left\vert f\right\rangle $. For an isotropic medium without
frequency dependence, this term is zero. However, it can become nonzero in
an anisotropic medium (e.g., $\varepsilon _{xx}\neq \varepsilon _{yy}\neq
\varepsilon _{zz}$). $\rho \digamma $ is similar as the spin torque dipole
density in spin transport theory \cite{Shi}, where the existence of the spin
torque reflects the fact that spin is not conserved microscopically in
systems with spin-orbit coupling. The term $\rho \digamma $ originates from
the dissipation of the medium, which induces the inconservation of the
density $\rho $ (i.e., $\dot{\rho}\neq 0$). We thus name $\rho \digamma $ as
the \textit{dissipation dipole density}.

\subsection{Degenerate bands}

In the above discussion, we consider only a single non-degenerate energy
band. The generalization of the theory to degenerate bands (e.g., left and
right circular polarization states of photons) is straightforward \cite{Niu4}%
. All above formulas are still valid by replacing the local wavefunction $f$
with $\sum_{i}\chi _{\alpha }f_{\alpha }$, where $f_{\alpha }$ is the Bloch
wavefunction at band $\alpha $ and $\chi _{\alpha }$ is the superposition
coefficient that can be determined through
\begin{equation}
i\dot{\chi}_{\alpha }=\left[ \left\langle f_{\alpha }\right\vert \mathcal{H}%
\left\vert f_{\beta }\right\rangle -i\left\langle f_{\alpha }\right\vert
\Theta _{0}\left\vert \frac{df_{\beta }}{dt}\right\rangle \right] \chi
_{\beta }.  \label{DB}
\end{equation}%
In a uniform medium without frequency dependence, Eq. (\ref{DB}) can be
simplified as $\dot{\chi}_{\alpha }=-i\mathbf{\dot{p}\cdot A}_{\mathbf{p}%
}\chi _{\beta }$ on a helical basis, which is the same as that derived in
\cite{onoda} using wavepacket dynamics. Eqs. (\ref{current3},\ref{diss},\ref%
{DB}) and the equation for $\mathbf{\dot{p}}$ are the new set of the EOM for
light propagation in photonic crystals.

\begin{figure}[t]
\includegraphics[width=0.7\linewidth]{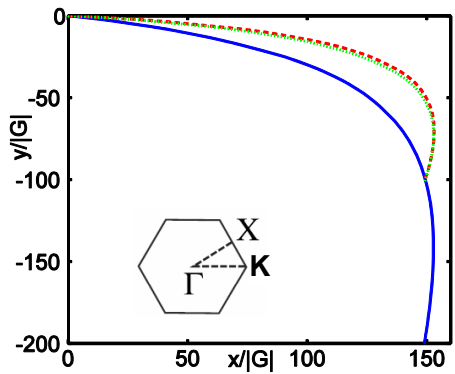}
\caption{(Color online) Plot of the trajectories of a photon in a 2D ($xy$)
hexagonal photonic crystal obtained from the geometric optics approach. The
length unit is $|G|^{-1}=a/2\protect\pi $, where $a$ is the lattice
constant. $\protect\lambda =0.5$, $\protect\eta _{0}=0.008$, $d\protect\eta %
_{1}/dx=4\times 10^{-6}$ with $\protect\eta _{1}\left( t=0\right) =0$. The
inital wavevector of the photon has $\mathbf{p}=\left( \Delta /3,0\right) $
from the BZ corner $\mathbf{K}$. Solid line: real trajectory of the photon.
Dashed line: trajectory without $\partial _{\mathbf{R}}\times \protect\rho
\mathbf{m}$. Dotted line: trajectory without both $\partial _{\mathbf{R}%
}\times \protect\rho \mathbf{m}$ and $\Omega _{\mathbf{pR}}\mathbf{\dot{R}}$.
}
\label{fig:trajector}
\end{figure}

\section{Transverse energy current from POAM}

Finally, we show that the additional energy current $\partial _{\mathbf{R}%
}\times \rho \mathbf{m}$ in $\mathbf{J}$ can be nonzero and may
significantly modify the photon transport in photonic crystals using a
concrete example. Consider a 2D hexagonal photonic crystal with a uniform
isotropic permeability tensor $\mu _{0}\delta _{ij}$, and an isotropic but
spatially-varying permittivity tensor $\varepsilon \left( \mathbf{r}\right)
\varepsilon _{0}$, with \cite{Haldane}
\begin{eqnarray}
\varepsilon _{ii}\left( \mathbf{r}\right) &=&\varepsilon \left( 1+\varsigma
V_{\mathbf{G}}\left( \mathbf{r}\right) \right) ,  \label{die} \\
\varepsilon _{xy}\left( \mathbf{r}\right) &=&-\varepsilon _{yx}\left(
\mathbf{r}\right) =-i\varepsilon \left( \eta _{0}+\eta _{1}V_{\mathbf{G}%
}\left( \mathbf{r}\right) \right) .  \nonumber
\end{eqnarray}%
Here $V_{\mathbf{G}}\left( \mathbf{r}\right) =2\sum_{n=1}^{3}\cos \left(
\mathbf{G}_{n}\cdot \mathbf{r}\right) $, $\mathbf{G}_{n}$ are the three
equal length reciprocal vectors in the \textit{xy} plane, rotated 120$%
^{\circ }$ relative to each other. The $\varepsilon _{xy}\left( \mathbf{r}%
\right) $ are the Faraday terms which explicitly break the time-reversal
symmetry and lead to nonzero Berry phase contribution to the photon
transport.

We focus on the decoupled TE set $\{E_{x},E_{y},H_{z}\}$ of the
electromagnetic fields. Around the corner K of the first BZ (see the inset
of Fig. \ref{fig:trajector}), the photon dynamics are governed by an
effective Hamiltonian \cite{Haldane}%
\begin{equation}
H=\omega _{0}I+c\left( p_{x}\sigma _{x}+p_{y}\sigma _{y}+\Delta \sigma
_{z}\right) /2,  \label{Hampc}
\end{equation}%
as shown in Ref. \cite{Haldane}, where $\omega _{0}=c\left\vert \mathbf{G}%
\right\vert \left( 1-\varsigma /4\right) /\sqrt{3}$, $\Delta =\left\vert
\mathbf{G}\right\vert \left( \frac{3}{2}\eta _{1}-3\varsigma \eta
_{0}\right) /\sqrt{3}$, $\mathbf{p}$ is the deviation of the 2D Bloch vector
from the BZ corner K, and $c$ is the speed of light. From the Hamiltonian,
we find
\begin{eqnarray}
\Omega _{z} &=&\pm \frac{1}{2}\Delta \left( p^{2}+\Delta ^{2}\right) ^{-3/2},
\label{Berry5} \\
\mathbf{m}_{z} &=&\pm \frac{c}{4}\Delta \left( p^{2}+\Delta ^{2}\right)
^{-1},
\end{eqnarray}%
where $\pm $ correspond to two photonic bands.

We assume the parameter $\eta _{1}\left( x\right) $ is slowly varying along
the $x$ direction. Consider a photon frequency at the lower band, a nonzero $%
\Omega _{z}$ leads to a transverse energy current
\begin{eqnarray}
\mathbf{J}_{\perp } &\mathbf{=}&-\Omega _{\mathbf{pp}}\mathbf{\dot{p}=}-%
\mathbf{\dot{p}\times \Omega }  \nonumber \\
&=&-\frac{\sqrt{3}c\left\vert G\right\vert \Delta ^{2}}{8\left( p^{2}+\Delta
^{2}\right) ^{2}}\frac{d\eta _{1}}{dx}\mathbf{e}_{y}  \label{current5}
\end{eqnarray}%
along the $y$ direction. However, the nonzero $\mathbf{m}$ also yields
another transverse energy current
\begin{equation}
\mathbf{J}_{\perp }^{\prime }=\partial _{\mathbf{R}}\times \mathbf{m}=\frac{%
\sqrt{3}c\left\vert G\right\vert \left( p^{2}-\Delta ^{2}\right) }{8\left(
p^{2}+\Delta ^{2}\right) ^{2}}\frac{d\eta _{1}}{dx}\mathbf{e}_{y},
\label{current6}
\end{equation}%
which is absent in previous literature. Note that these two currents have
the same directions around $\mathbf{p}=0$ and are comparable in magnitudes.
In Fig. \ref{fig:trajector}, we plot the trajectories of a photon with and
without taking account of the contribution (\ref{current6}) from the POAM.
We see the clear distinction between these two trajectories, which indicates
that the transverse energy current from the POAM cannot be neglected in the
transport of photons in photonic crystals. Finally we emphasize that the
photon trajectories in Fig. 1 are obtained from our first order geometric
optics treatment and it would be interesting to compare them with the fully
correct photon trajectories obtained from ab initio calculation in the
future.

\section{Conclusion}

In summary, we derive a new set of semiclassical EOM of particles in a
chiral and dissipative periodic medium subject to slowly varying
perturbations in space and time. The effects of several important properties
of particles (or waves) and media that are neglected in current literature (%
\textit{e.g.}, self rotation of particles, frequency (or energy) dependence
and dissipation of the media) have been included in the EOM. The EOM are
derived based on a geometric optics approach for light propagation in
periodic photonic crystals. We show that Berry curvatures in position, time,
momentum, and frequency spaces, the dissipation dipole density of the
medium, and the photon angular momentum may affect the Poynting vector field
of photons that traces the geometric optics rays. Reciprocal as well as
effective electromagnetic fields for photons may arise from the Berry phase
effects in photonic crystals. We emphasize that although the geometric
optics theory\ and the EOM are presented for the light propagation in
photonic crystals, they can also be applied to many other systems, to name a
few \cite{Nagaosa1}, electron dynamics in Bloch bands, X-ray propagation in
natural solid crystals that can be taken as photonic crystals for X-rays,
light propagation in multiferroic materials, quasiparticle dynamics in a
superfluid, and transport of acoustic waves.

\textbf{Acknowledgement:} Zhang is supported by WSU Startup. Niu is
supported by the NSF, DOE, R. A. Welch foundation (F-1255), and Texas
Advanced Research Program.

\end{document}